\documentclass[conference]{IEEEtran}
\IEEEoverridecommandlockouts
\usepackage{cite}
\usepackage{amsmath,amssymb,amsfonts}
\usepackage{algorithmic}
\usepackage{graphicx}
\usepackage{textcomp}
\usepackage{hyperref}

\def\BibTeX{{\rm B\kern-.05em{\sc i\kern-.025em b}\kern-.08em
    T\kern-.1667em\lower.7ex\hbox{E}\kern-.125emX}}

\AtBeginDocument{
  \label{CorrectFirstPageLabel}
  
}	
	
\begin{document}

\title{Understanding Social Factors Affecting The Cryptocurrency Market}

\author{\IEEEauthorblockN{1\textsuperscript{st} Gourang Aggarwal}
\IEEEauthorblockA{\textit{Computer Science Department} \\
\textit{NIIT University, Nemrana}\\
Alwar, Rajasthan, India \\
gourang.aggarwal@st.niituniversity.in}
\and
\IEEEauthorblockN{2\textsuperscript{nd} Vimal Patel}
\IEEEauthorblockA{\textit{Computer Science Department} \\
\textit{NIIT University, Nemrana}\\
Alwar, Rajasthan, India \\
vimal.patel@st.niituniversity.in}
\and
\IEEEauthorblockN{3\textsuperscript{rd} Gaurav Varshney}
\IEEEauthorblockA{\textit{Computer Science Department} \\
\textit{NIIT University, Nemrana}\\
Alwar, Rajasthan, India \\
gaurav.varshney@niituniversity.in}
\and
\IEEEauthorblockN{4\textsuperscript{th} Kimberly Oostman}
\IEEEauthorblockA{\textit{Department of Communication and Journalism} \\
\textit{University of New Mexico Albuquerque}\\
New Mexico, USA \\
kroostman@unm.edu}

}

\maketitle

\begin{abstract}

Blockchain and its application on cryptocurrency transactions have gathered a lot of attention and popularity since the birth of the pioneer Bitcoin in 2009. More than 1500 cryptocurrencies are currently circulated in the market. The technology underpinning Bitcoin and other cryptocurrencies is Blockchain and is a rapidly growing decentralized distributed-ledger technology which find its major involvement in cryptocurrencies. But cryptocurrencies are of extremely volatile and fragile nature which makes it difficult to be used as a stable currency for transactions and devoid this market of human trust. Cryptocurrency market is controlled by various social and government factors which keeps it fluctuating. This paper identifies and discusses the important factors that govern the cryptocurrency market and analyzes the impact of these factors. A pilot user survey has also been presented at the end of this paper to understand and demonstrate the societal view of the acceptance of cryptocurrencies.
\end{abstract}

\section{Introduction}
The cryptocurrency market has evolved erratically and at unprecedented speed since Bitcoin came into existence. The underlying technology supporting cryptocurrencies, the blockchain, is evolving at a rapid speed and financial institutions such as Accenture, JP Morgan, UBS, and others are joining together with other big companies to form an alliance and improve blockchain to make it faster, more secure and reliable~\cite{ref_book2}. A cryptocurrency is a form of electronic cash that relies on cryptography. It functions much like a standard currency, enabling users to exchange virtual payments for goods and services, which does not require a central trusted authority. This reduces the exploitation of human inefficiencies (middle-man transaction cost, credit card, and debit card transaction commission and the inability to sell/buy commodities directly). Segments of societies from all over the world are swiftly moving towards the cryptocurrency and its usage despite its nonacceptance from the governments of major influential countries such as India, South Korea, the United States and Japan. People from various countries with libertarian ideals are attracted towards the rebellious attribute of the cryptocurrency i.e. its decentralized nature~\cite{ref_book37}. This method requires users to trust the blockchain system more than to trust a middle party or organization with their data. This aspect is one of the major characteristics of cryptocurrencies which made them very popular and highly acceptable. At the time this paper was written, the cryptocurrency industry consisted of over 1500 cryptocurrencies with a market capitalization of over 430 Billion USD~\cite{ref_book3}. Due to the volatility of this market, the capitalization of cryptocurrency changes frequently daily often by a large amount.

\begin{figure*}
    \centering
    \includegraphics[width=0.8\textwidth]{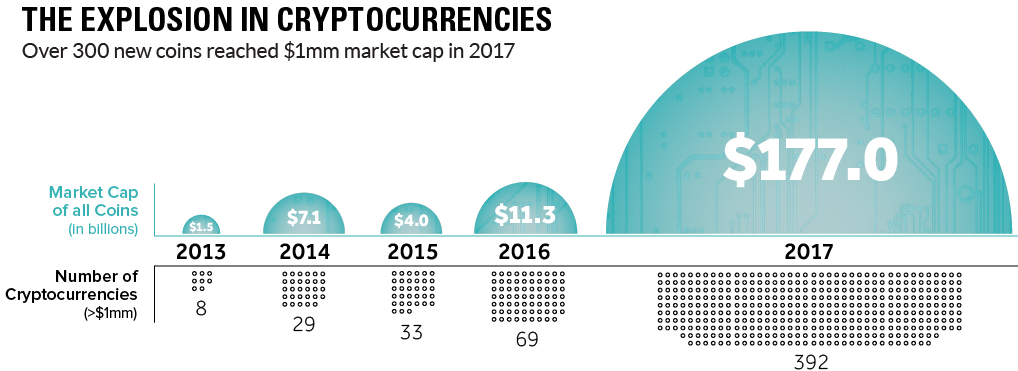}
    \caption{Increasing popularity of cryptocurrency~\cite{ref_book4,ref_book38}}
    \label{fig1}
\end{figure*}
Figure~\ref{fig1} depicts how cryptocurrency has gained a lot of attention especially from 2014 to 2018. The peaks and drops are visible at almost every point in the Figure~\ref{fig2} which is a clear indicator of the volatility of the cryptocurrency market. Some of the reasons which could influence the growth and price of this cryptocurrency market are identified by researchers such as government regulations and personal perceptions~\cite{ref_book5}. However, there could be important factors that may be affecting the market price and fluctuations.
After the review of many papers, the authors observed that no current study clearly describes the potential factors and their impact. This motivated the authors to dig dipper into the crypto-world to collect some data that may indicate these factors and their hand in the pricing field.
The main contribution of this paper is as follows:
\begin{enumerate}

    \item Authors performed an initial analysis of the problem of high volatility of the cryptocurrency market and its dependence on various social factors via a real-time study on the data collected for a period of six months.
    \item Authors identified a set of factors which affect the cryptocurrency market based on theoretical inputs and real-time market data.
    \item Authors provide the correlation values of the identified factors with the rise or fall of cryptocurrency market using data mining tool Weka.
\end{enumerate}
The paper is organized as follows: Section 1 of the paper provides a brief introduction to the world of cryptocurrency. Section 2 provides a literature review exploring the reasons behind the fluctuating market of cryptocurrency and the possible important attributes which the authors believe were not observed by other researchers. Section 3 examines the various attributes and how they affect the price by performing data mining using various tools and the techniques. The section presents a thorough analysis of these factors to gain knowledge. Section 4 concludes the paper and discusses future research directions.

\section{Literature Review}
Section 2 provides a literature review which, first, discusses current studies associated with crypto-market fluctuations. Next, the media effects communication theories of agenda-setting and framing are defined. Finally, a justification for further examination of factors affecting cryptocurrency market volatility is provided.

\subsection{The Cryptocurrency Market}
The cryptocurrency market is highly volatile and the price fluctuates dramatically. The exact reasons for the large variations in cryptocurrency prices are mostly unknown and this has become a challenging problem to be explored by computer and social scientists. Some factors that might affect the price of cryptocurrency identified during previous research include: the non-intrinsic value of cryptocurrencies, as they mostly depend upon market sentiments, lack of regulatory oversight, since the crypto-market is so free and unregulated that causes high market manipulation by some key players who hold a large volume of the overall cryptocurrency, herd mentality (thousands investing and divesting in fear of losing money)~\cite{ref_book1}.

These factors as identified by previous researchers might have a potential hand in the volatility of the cryptocurrency market. However, there are relatively few studies on the topic and therefore a lack of validation of the operational factors and their effects on the cryptocurrency market. Even if we name the identified factor and assume them to be correlated with cryptocurrency fluctuations, there still could be a lack of some other social factors from multiple sources which might be impacting the volatility of the cryptocurrency market. Li and Wang had suggested that there could be factors that are related to social media sentiments which in turn might be correlated to cryptocurrency volatility~\cite{ref_book6}.

Balcilar, Bouri, Gupta, Roubaud determined that trading volume can predict returns but does not affect the price~\cite{ref_book7}. Hayes found that relative differences in costs of production on the margin are the main determinants~\cite{ref_book8}. However, these studies did not cover the whole cryptocurrency market, the only bitcoin with a few reasonable factors, leaving behind a scope to identify and room to discover other crucial social factors.

\subsection{Media effects}
An important consideration in the interplay of social factors affecting the cryptocurrency market is the effects of media on public opinion.  Within mass media communication, newspapers, radio, television, and now web and mobile phone technologies, are all means to convey information to the public. Mass communication agenda-setting and framing theories focus on the media’s ability to tell the public what to think about and how to think about salient issues — perhaps also what to do about them. Tsvetkova, Yasserri, et al. examined human-machine relationships interacting with prediction markets and determined that information and trust are key factors impacting market capitalization and user confidence in these exchanges (2017)~\cite{ref_book34}.

\subsection{Agenda-setting}
Mass communication scholars have long studied media effects on society. Media effects are described as “the social, cultural, and psychological impact of communicating via the mass media”~\cite{ref_book27}. In 1972, McCombs and Shaw introduced the mass communication agenda-setting concept to explain the phenomena of news sources imposing the salience of stories during political campaigns, which in turn affected voters’ priorities~\cite{ref_book36}. In other words, the topics that voters considered most important were those that had received the most news coverage. In essence, it was determined that media reports of news stories, exposure parameters, and frequencies create some effects on audience members within the realm of political communication.

\subsection{Framing}
The framing of prominent issues also impacts media consumers. Along with media telling citizens what to think about via agenda-setting effects, framing assumes that how information is characterized, can affect the understanding by, and influence over, an audience (Scheufele and Tewksbury, 2007)~\cite{ref_book35}. Entman explained the concept of framing as the “way in which influence over a human consciousness is exerted by the transfer (or communication) of information from one location such as a speech, utterance, news report, or novel-to that consciousness (Entman, 1993, pp. 51-52)~\cite{ref_book30}. Further, Scheufele and Tewksbury (2007) explicate framing as a macro-construct whereby journalists and other sources take care to present news in a relatable way to their audience and the audience’s existing schema, as well as a micro-construct in which information is received and the presentation features are used by an individual to understand and form impressions~\cite{ref_book35}. In order to make information more prominent, frames are constantly generated. Framing helps to make material credible, easier to understand, and relevant. In order to accomplish successful audience perception, frames either highlight or omit information. Effective framing takes into consideration audience schemata to guide closely related ideas~\cite{ref_book29}. In this way, framing media effects impact audience members.

Other media scholars also made assertions about media effects. Lang and Lang (1966)~\cite{ref_book33} claimed that mass media present information that individuals should “think about, know about, have feelings about” (p. 468) while Cohen~\cite{ref_book28} stated that the press “may not be successful much of the time in telling people what to think, but it is stunningly successful in telling its readers what to think about” (1963, p. 13). Since the introduction of agenda-setting and framing theories in mass communication which describe ways that public determine salience of information presented, and form opinions from that information, hundreds of articles and studies have examined the evolution and utility of the theories. Most studies investigate these media effects in the political communication process (McCombs and Shaw, 1972; Iyengar, Peters, and Kinder, 1982; Iyengar and Kinder, 2010)~\cite{ref_book31,ref_book32,ref_book36}. Therefore, over the last several decades, empirical evidence supports media agenda-setting and framing effects.

\subsection{Social Factors}
In the various literature reviewed, we identified a set of social factors which we believe to have a role in the price fluctuations of cryptocurrencies. These factors were shortlisted based on the prior knowledge obtained from previous research work and analysis of important events in the cryptocurrency market which were recorded over a period of six months. Finally to verify the role of these factors we collected the needed data from various Internet sources such as news and other social media or blog articles published on the same day or a day before during a sharp peak or drop in the popular cryptocurrency prices. The factors which we believe to be correlated and further analyzed in our study include: banks notices regarding cryptocurrency, news from cryptocurrency exchanges, government regulations, opinions of known tech personalities having relation with crypto-industry, views of popular celebrities, information that relates to cryptocurrency bans, announcements regarding cryptocurrency from big companies or regulatory authorities and information of a cryptocurrency fraud which are further divided based on whether the inputs coming from them are considered positive towards cryptocurrency (supporting the price positively) or negative.

\subsection{Summary}
After studying previous researches and analyzing various sources of information we believe that researchers may have missed out analyzing/studying the role of some important social factors which might be affecting the prices of cryptocurrencies. In addition, we understand that media effects may provide opportunities to examine whether media sources or topics that could impact cryptocurrency market volatility.

\section{Identifying Factors Affecting Cryptocurrency Market}
In the previous section, we listed the various social factors from our preliminary analysis. In this section, we will do further analysis of these factors and will infer whether these factors are actually affecting the prices or not.

\subsection{Tools and Techniques}
To perform this study we have used the following tools and techniques:
\begin{itemize}
    
    \item {Data Collection} has been performed via manual analysis of cryptocurrency data for six months by two individuals reading articles over the internet. Google News regarding cryptocurrencies from all major countries including the USA, India, Australia, EU etc. was read daily. Articles from Coindesk, Coinmarketcap, and Coinbase were followed as well as twitter post with cryptocurrency related hashtags were monitored on a daily basis.
    \item {Attribute/Factor Identification} has been done by the authors for the very first time as no substantial concrete text was available on factors influencing the cryptocurrency market. The authors used human physiological traits, social media factors integrated with inputs from the historical cryptocurrency data received during data collection to come up with a set of factors. All the factors identified were given suitable names. 
    \item{Weighing and Scaling-} All the attributes in are given a value based on its occurrence whether that occurrence had an increasing or decreasing impact on the price of cryptocurrency. They tried to see if the attributed persisted during a peak or fall of the prices of cryptocurrency and marked it accordingly using a predefined convention discussed in detail in the following section \ref{weighing}. The markings were then discussed and concluded the authors.
    \item Correlation analysis has been done to find the important attributes which have a higher correlation with rise or fall of the cryptocurrency market. Authors opted for Weka machine learning tool.     
\end{itemize}

\subsection{Data Collection}
We first recorded the dates at which major peak and drop happened over the last six months. By peak, we mean a sudden increase in cryptocurrency prices and a drop means a sudden decline in the cryptocurrency prices. Data of top news stories were collected for a day before a major peak or drop came, assuming that the news had affected the peak or drop that happened the next day. We collected all the top eight to fifteen incidents and events that happened during the X-1 day via news/article/blogs/twitter posts etc for each major peak on that day sorted by relevance which seemed to have an impact on the change of the price of the cryptocurrency market.
\begin{figure*}
    \centering
    \includegraphics[width=0.99\textwidth]{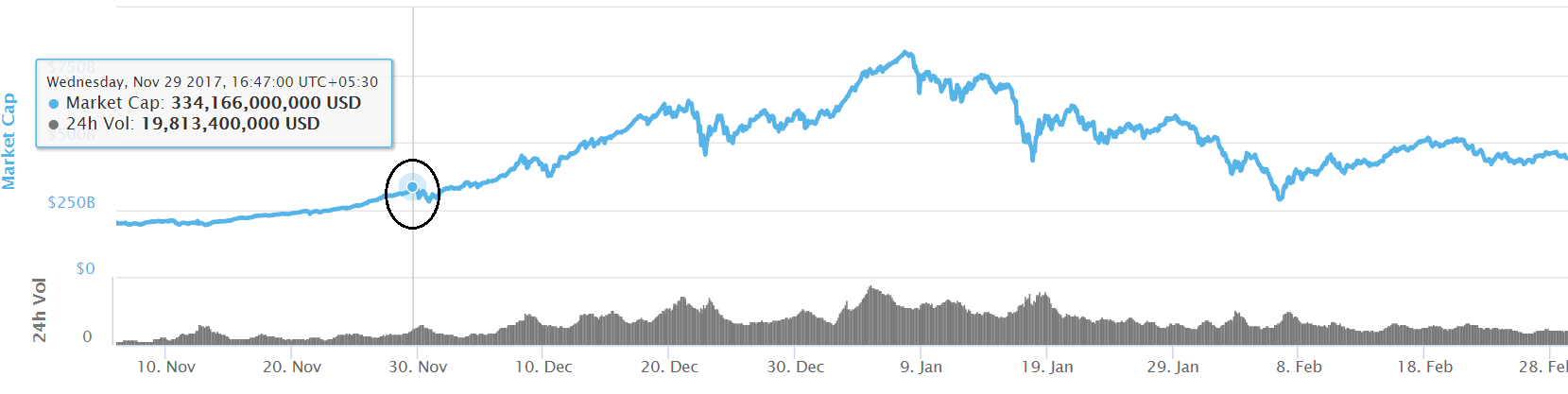}
    \caption{Data selection based on peaks and drops~\cite{ref_book4}}
    \label{fig2}
\end{figure*} 

For example, from Figure~\ref{fig2} one can see that the first major drop for the whole cryptocurrency market came on 29 Nov 2017, when the market capitalization rose over 330 billion USD~\cite{ref_book4,} to drop to 278 billion USD on 30 Nov 2017. It was the first major drop in the data collected over four months as the black circle depicts in  Figure~\ref{fig2}.

\subsection{Attribute Identification}
All the attributes that were identified in the study were chosen on the basis of the fact that such announcements and views have a tendency to influence the attention of the crowd. For example, celebrities possess very high influential power. The followers can change their perception about cryptocurrency based on the celebrity's viewpoints and/or thinking. Whatever they say whether right or wrong is strongly accepted by their fans. One article such as "How Do Celebrities Really Affect Us?" discusses the power celebrities have to change the perception of a person~\cite{ref_book9}.

\begin{figure*}
    \centering
    \includegraphics[width=0.95\textwidth]{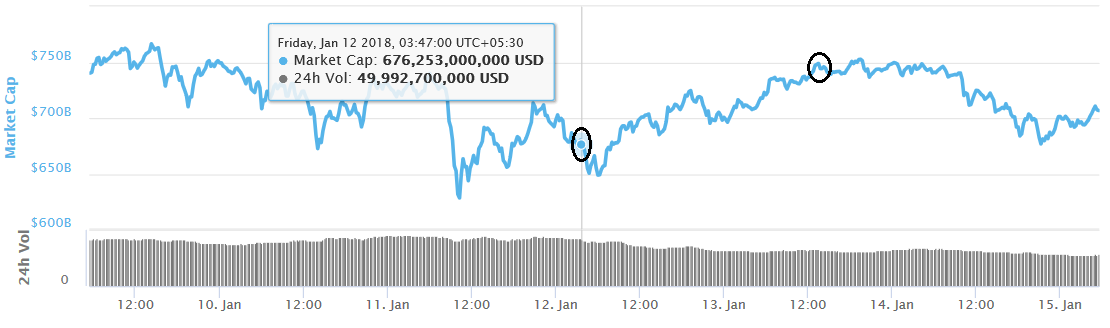}
    \caption{Price on 12 Jan and 13 Jan are shown with circles~\cite{ref_book4}}
    \label{fig3}
\end{figure*}

Similarly, for the attribute \texttt{Banks Notices Regarding Cryptocurrency}, we believe that any notice that comes from the banks especially from the central banks can mold the public ideology. For example RBI, the central bank of India regularly passed notices during the last six months to warn people about cryptocurrencies being a Ponzi scheme~\cite{ref_book10}. As explained by Entman's framing theory, when a bank of such a high reputation releases its views on the financial topic, the public mindset could be influenced. This may result in creating negative views on the usage of cryptocurrencies~\cite{ref_book30}. Similarly, \texttt{Information that relates to Cryptocurrency Bans} might impact the mindset of the people. The impact can be either positive/negative or neutral depending on the nature of the news. This scenario can be observed through the example of South Korea where the government backed off from banning the trade of cryptocurrencies~\cite{ref_bookl1}. This news came on 12 Jan 2018 when the cryptocurrency market was struggling. Immediately, on 13 Jan 2018 after this news report, a positive impact was noted on the market price, and the price continued to rise, as compared to the price on 12 Jan 2018. Figure~\ref{fig3} shows the price on 12 Jan 2018 and 13 Jan 2018 respectively, represented by the first and second circle (the leftmost being the first). This news story can be considered as one of the social factors that might have affected the price. In our analysis over a period of six months, we identified other social factors which could have created a potential positive or negative impact on the news recipients. While many white papers were consulted for writing this paper, the majority of knowledge comes from synthesizing the raw data available from Internet sources.

\subsection{Weighing and Scaling Attributes} \label{weighing}

Each attribute was divided into two sides: the positive and the negative one, the ban and the fraud attributes were given only a negative score since ban and fraud are themselves negative and cannot have a positive effect. The final class label attribute can only contain two values either a peak or a drop. Every news report, obtained from input sources were given either 0 or 1 taking the following points into account: 

If the nature of the news is against cryptocurrency(i.e negative), then the positive column of all the attributes automatically becomes 0 and the points to the negative column of individual attributes are given on this basis:\texttt{0}: If the news does not contain or has any connection or relation with the attribute and \texttt{1}: If the news does contain or has any connection or relation with the attribute.
If the nature of the news is positive, then the negative column of all the attributes becomes 0, and the points to the positive column of an individual attribute are given on the same basis written above. For the Ban and Fraud attribute this criterion was used: \texttt{0}: If the news does not contain or has any connection or relation with the attribute and \texttt{1}: If the news does contain or has any connection or relation with the attribute.

\begin{figure*}
    \centering
    \includegraphics[width=0.95\textwidth]{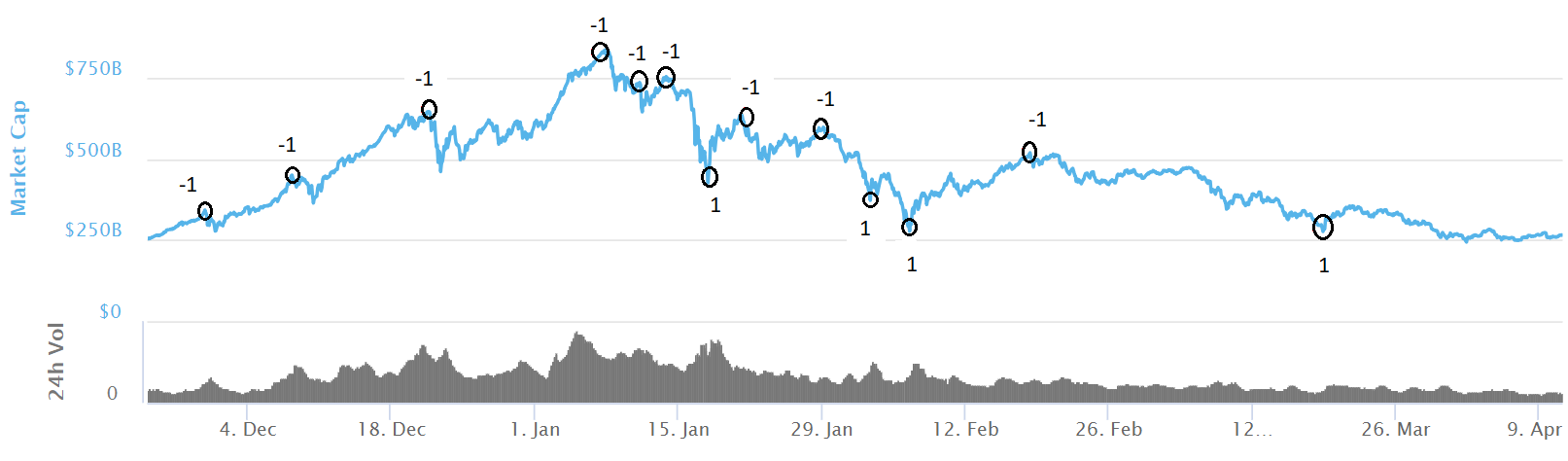}
    \caption{Peaks and Drops considered for the data set used for the study~\cite{ref_book4}}
    \label{fig4}
\end{figure*}

For example- a news story "Bankers Have Shut Down All of Bulgaria’s Bitcoin Exchanges" ~\cite{ref_book12}the following are negative news for cryptocurrencies, so only negative part of the attributes are given the point as 1 i.e. \texttt{Banks} \texttt{Notice/}\texttt{News} \texttt{Regarding} \texttt{bank}\texttt{(negative)} and \texttt{Bans}.
Since the news talks about Bulgaria’s major banks shutting down major cryptocurrency exchanges. The news story is negative for the cryptocurrency market. This news is about banks hence the first attribute which got its value as 1 is Banks Notice/News Regarding bank and only the negative attribute got its value as 1 since its a negative news story from the banks. Secondly, the news is related to the ban of the cryptocurrency exchanges hence the attribute Bans(negative) or simply Bans also derived its value as 1 point. All other attributes do not seem to fall into the news story and were assessed a value of 0.
 
 \subsection{Data Pre-Processing, Sentiment Analysis}
Once the data was well populated, another column was created with a name meta-data, which basically consisted of a summary of that news. A summarizing tool was used to create a summary of each news article we obtained from the Internet~\cite{ref_book13}. Another tool named Lexalytics was used to do phrase sentiment analysis~\cite{ref_book14}. Phrase sentiment analysis gives a value between -1 and 1 as follows: \texttt{-1}: If the phrases are portraying negative sentiments and towards \texttt{1}: If the phrases are portraying positive sentiments. Next, all the values of the other attributes which we recorded in the earlier section were multiplied with the corresponding result obtained for each row(news) from the sentiment analysis. A serial number was set for all the dates. For example, 29-Nov-17 was given a serial no. 1,08-Dec-17 was given a serial no. 2 and so on. Then, all the news recorded for each date was inserted into a pivot table with serial no. as a single row and all the data of an attribute was added in each column for a specific date and used as a single value.

\begin{figure*}
    \centering
    \includegraphics[width=0.99\textwidth]{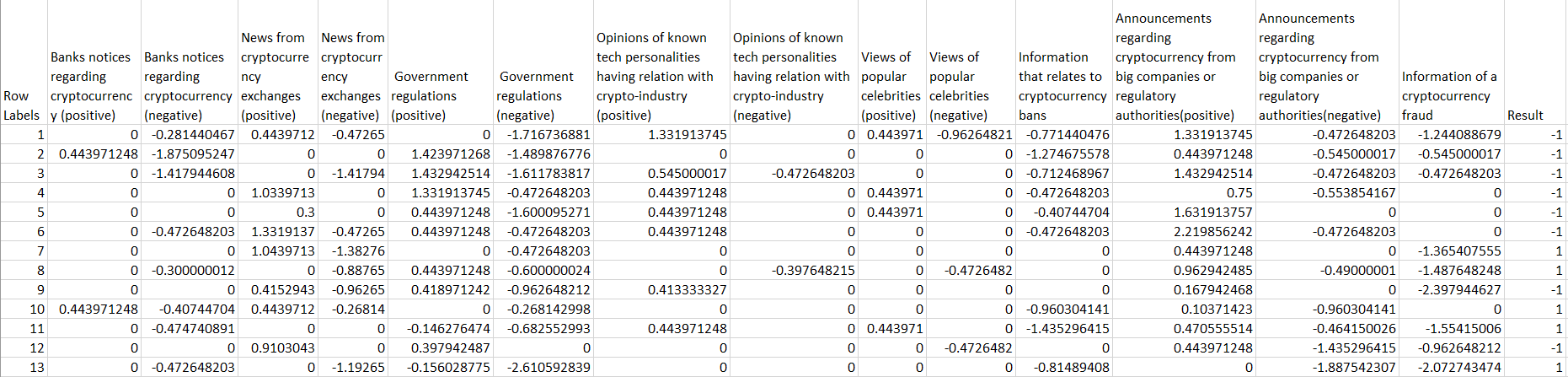}
    \caption{Final data imported into the tool}
    \label{fig5}
\end{figure*}
In the end, a result column (class label attribute) was added and values were filled corresponding to whether the market performed well or went down in those dates, i.e. drop(-1) and peak(1). The pivot table, as you can see in the Figure~\ref{fig5} was created as the final data, which was saved as a .csv file(Comma Delimited) to be used for analysis.
The original data can be referred to through the following link:(uploaded and shared on Google drive):  
\href{https://docs.google.com/spreadsheets/d/1XBUf9wD-MLMK_HccbdGXy3LSQnWQfc83LqdrgKeiqF4/edit?usp=sharing}{Dataset}

\subsection{Important factors affecting the price} ~\label{sec3.5}
The data mining tool Weka was used to identify the correlation between all the attributes collected in earlier steps and their effect on the cryptocurrency fluctuations/prices which were set as the result attribute~\cite{ref_book15}. The result attribute was the class attribute against which all the other attributes were evaluated. The feature/attribute selection was divided into two parts~\cite{ref_book16}: Attribute Evaluation and Search method. Our main aim was to find the most important factors which might affect the price of the cryptocurrency. To achieve this we used CorrelationAttributeEval as the Attribute Evaluator to find the correlation analysis between the class attribute and other attributes~\cite{ref_book17}. This method evaluates the worth of an attribute by measuring the correlation (Pearson's correlation~\cite{ref_book18}) between it and the class. We used Ranker-T method in the search method which was used to choose the attributes in order to evaluate the class attribute~\cite{ref_book26}.
Each attribute was evaluated against the class attribute(result attribute) and the results were obtained which can be seen in Figure~\ref{fig6}:

\begin{figure*}
    \centering
    \includegraphics[width=0.95\textwidth]{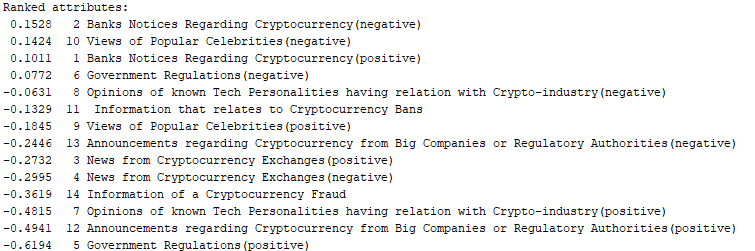}
    \caption{Calculated results through Weka}
    \label{fig6}
\end{figure*}

The results that are inline with the information recipient's beliefs in real practice are Banks Notices Regarding cryptocurrency(positive)(3rd from the top) which has a correlation value of 0.1011 (weak) , it is inline with the belief that a positive news from a bank can actually lead to an increase in market price. Similarly, the 5th attribute i.e. Opinions of Known Tech Personalities Having Relation With Crypto-industry(negative) has correlation value -0.0631 which shows that a negative news story from a big personality or celebrity might drop the cryptocurrency price. Similarly, the Information that Relates to Cryptocurrency Bans(negative) (the 6th attribute), has a correlation with a value of -0.1329 can also lead towards a drop in the price of cryptocurrency. Any announcements regarding cryptocurrency from Big Companies or Regulatory Authorities(negative) (8th attribute), has the correlation value of -0.2446. A news story regarding Information of Cryptocurrency Fraud is negative news which makes people mistrust the use of cryptocurrency can be observed in Figure~\ref{fig6} also has a correlation of -0.3619.

But some results were contradictory to the normal (expected) belief or very unusual. For example, Government Regulations(positive)(1st from bottom) has a correlation value of -0.6194 which is contradictory to the belief that any positive news from the government should lead to a hike in price. It means that a positive news regarding Government regulations(positive) instead of leading to hike in price is leading to a drop in the market. Similarly for Government regulations(negative)(4th from the top) where it shows a correlation of 0.0772 although coming from a negative news story. The authors believe that this might be due to the fact that people interpret regulations in different ways. A series of positive and negative news from government can also contribute to these contradictory correlations. For example, South Korea presented a positive news story stating "South Korea Outlines Proposed Legislation for Cryptocurrency Exchanges"~\cite{ref_book19} but at the same time also reported a negative news stating "South Korea Leader Fears Bitcoin Leads Youth to Drugs"~\cite{ref_book20} which contradicted the two beliefs and might have created these conflicting correlations. However, we are still studying this factor further to come to a conclusion.

This contradiction may also affect news regarding the Opinions of Known Tech. Personalities having Relations with Crypto-industry. For example, a statement was released which stated: "Baroness Michelle Mone launches cryptocurrency to 'encourage women to invest in tech'"~\cite{ref_book21} which was a positive news story regarding cryptocurrency coming from a known personality Baroness Michelle Mone~\cite{ref_book22}. Around the same time, JPMorgan released a negative statement "JPMorgan misses the point, says crypto Isn't Currency"~\cite{ref_book23}. Both statements came from famous personalities and contradicted each other. We believe such contradictory news might have caused these conflicting results. Same ideology corresponds to Announcements Regarding Cryptocurrency from Big companies/Regulatory authorities. Also, there exists a very unusual attribute regarding News from Cryptocurrency Exchanges (Positive and Negative) (6th and 5th from the bottom), which were coming together with almost the same negative correlation of -0.2732 and -0.2995 respectively. All of these attributes which have a correlation between -0.2 and 0.2 can be ignored since they hold minimal significance. Hence, all the positively correlated result we achieved should be ignored since they are less than 0.2 and only the last 7 results should be considered. Out of these 7 results, 4 of the results obtained were contradictory to the normal or expected belief which includes News from Cryptocurrency Exchanges(positive), Opinions of Known Tech. Personalities having Relation with Crypto-industry(positive), Announcements Regarding Cryptocurrency from Big Companies/Regulatory Authorities(positive) and Government Regulations(positive).

There are several possible reasons for the contradictory/unusual results:
One possible reason for this kind of unexpected result might be the size of the data set. We collected data from the past 4 months and only collected top news stories for the major peaks and drops in cryptocurrency prices. To have a more definite analysis a larger and deeper data set is be required to show better and more accurate correlations. Secondly, in a few cases, there are news stories that are very specific to a single country, which will only affect the trading or usage of cryptocurrencies in that country. Sometimes the trading volume from that particular country might not be significant enough to give major fluctuations in the global market. We currently considered and weighted all news from all regions to be equal but a specific weight should be utilized and that should depend on the trading volume of the country whose news is getting considered. We believe this factor should have also been included in the research study and our ongoing work will include it in the future.

Along with the quantitative results we obtained, we can also infer that for the same reasons, it is difficult to determine any direct media effects from agenda-setting or framing of news stories. Although it is plausible that a positive for a negative presentation of a story, the frequency and distribution of coverage can impact what people think about cryptocurrency, and therefore its valuation, more data needs to be collected to draw clearer correlations between attributes and market results. Also, there are some news stories that would have affected the cryptocurrency more than any other news on the same day. This can be linked with most read or most circulated news on the web that day. There should have been a ranking method in the data to evaluate the weight of individual news articles considered based on the rank in which they are displayed at news.google.com or based on their number of reads or shares. For example, top news of that day could have been given rank one and subsequent news could have been given a lower rank, and then a proper weight could have been given to news based upon its ranking. Our future work is focused on including this as well. 

Some of the social factors which are identified to be correlated in Section~\ref{sec3.5} have been further validated using real time social media data which was collected from micro-blogging website Twitter using \texttt{tweepy} API. The first factor we analyzed was \texttt{Information} \texttt{that} \texttt{relates} \texttt{to} \texttt{Cryptocurrency} \texttt{Bans}. Recently a tweet saying that \textit{The Reserve Bank of India banned the transfer of bank account cash into bitcoin} from [Forbes] [14.87 M Followers] at 2:15 PM on 6th April 2018 caused no big change in the cryptocurrency prices that day. This validates the finding that \texttt{cryptocurrency} \texttt{bans} has not a very serious correlation with cryptocurrency prices. The second factor that we have validated is \texttt{Announcements} \texttt{regarding} \texttt{Cryptocurrency} \texttt{from} \texttt{Big} \texttt{Companies} \texttt{or} \texttt{Regulatory} \texttt{Authorities} \texttt{(positive)}. One of the recent tweet by \texttt{Ethereum} \texttt{King}[42.7 K followers] at 10:47 AM on 6th April \textit{SALE IS CURRENTLY LIVE with 30\% BONUS http:// goo.gl/yBxiBT ethereum investing altcoinnews cryptoinvesting gifcoin Profitable tokensale ROI ProfitMaker pic.twitter.com/Vsr11HXWwc}, caused a decrease in the Ethereum price during this time which somewhere validates a high negative correlation of -0.4941. By the time of writing this paper we were able to validate only some of the factors using the real time twitter dataset. This is an initial validation and we are working on a more rigorous validation of the results received in Section ~\ref{sec3.5}. 

\section{Conclusion}
From the initial research work presented in this paper, it was inferred that the correlation between the social factors and the market trends does not seem to hold a strong significance i.e there exists no strong correlation between the social factors and the market trends. The social factors that were analyzed were very similar to the way the stock market is analyzed~\cite{ref_book39,ref_book40,ref_book41}. This shows that the dynamics of this cryptocurrency market is very different to the way traders expect. Cryptocurrency is not a stock to invest in. It should not be analyzed the way the stock market is analyzed. The volatility of the market is very dynamic and it is very hard to predict which factor might affect how much at any point in time since cryptocurrencies are not blocked within borders. Although the different news stories come from a variety of sources and are presented to appeal to the existing schema of an audience, it is still too early in the emergence of cryptocurrency and due to our limited data collection to find a direct media effect between news reports and market fluctuation. At any specific time, various positive and negative news of different kind related to cryptocurrency are coming out, which might have short term or long term effect. Hence, it becomes very hard to give weight to any factor since a corresponding opposite kind of news might also come up at that very time to nullify the effect of that news. 

However, deeper work is required to be done in this field. A larger dataset is required to get better and more reliable results. Other issues must be considered, as the inclusion of the ranking of news and the inclusion of the trading volumes of the country from where the news is originating. These factors should be weighted in the data.

\end{document}